\documentclass[a4paper,12 pt]{article}
\usepackage{amsfonts}
\usepackage{eufrak}
\usepackage{amsthm}
\usepackage{amscd}
\usepackage{ulem}
\usepackage{cancel}
\usepackage{amssymb}
\usepackage{graphicx}
\usepackage{color}
\usepackage{graphics}
\usepackage{setspace}
\usepackage{anyfontsize}

\newcommand{\tp}{t_p}

\newtheorem{lem}{Lemma}

\theoremstyle{definition}

\theoremstyle{remark}

\newcommand{\dst} {\displaystyle}

\newcommand{\la} {\lambda}

\newcommand{\veps} {\varepsilon}

\newcommand{\lb} {\left(}
\newcommand{\rb} {\right)}
\newcommand{\lbr} {\left\{}
\newcommand{\rbr} {\right\}}
\newcommand{\ls} {\left[}
\newcommand{\rs} {\right]}

\newcommand{\ld} {\left.}
\newcommand{\rd} {\right.}
\newcommand{\lv} {\left|}
\newcommand{\rv} {\right|}

\newcommand{\ga} {\gamma}
\newcommand{\bga}{\begin{array}{l}}
\newcommand{\ena}{\end{array}}
\newcommand{\bge}{\begin{equation}}
\newcommand{\bgea}{\begin{equation} \begin{array}{l} }
\newcommand{\ene}{\end{equation}}
\newcommand{\enea}{ \end{array} \end{equation}}

\newcommand{\text}{ }

\DeclareMathAlphabet{\pazocal}{OMS}{zplm}{m}{n}

\begin{document}
\author{Maxim Zyskin}
\title{Single Particle Battery Model via Universal Transform Method}

\date{}
\maketitle

\begin{abstract}
We obtain and investigate an explicit solution via universal transform  of the diffusion equation in a spherical particle which appears in the so-called single particle model, a popular simple model of an electric battery.
\end{abstract}

\renewcommand{\tp}{t_+}
\newcommand{\tpg}{\tp}
\newcommand{\gatp}{\ga_{+}}
\newcommand{\gai}{\ga_i}
\newcommand{\gal}{\ga_{0}}
\newcommand{\gar}{\ga_1}
\newcommand{\ui}{u_0}
\newcommand{\hui}{\hat{u}_0}
\newcommand{\huim}{\hat{u}_{0m}}
\newcommand{\huip}{\hat{u}_{0p}}
\renewcommand{\ni}{n_0}
\newcommand{\Dp}{D_+}
\newcommand{\Dm}{D_-}

\section{Introduction}
In recent years significant efforts were attracted to electric battery research \cite{Goodenough2013} and modeling \cite{Newman2021}, due to the goals of transition to cleaner energy, and the need to design better electric batteries to meet such goals.

In the simplest electric battery model, the so-called single particle model \cite{rev1}, one has to solve for concentration of ions in a single electrode particle, satisfying a diffusion equation, given the spherically symmetric flux of ions at the surface of the particle. Concentration of ions at the surface of the electrode determines electrochemical potential, according to a chemical reaction rate law, such as Butler-Volmer equation. Such single electrode particles, one for anode and another for the cathode, represent all respective electrode particles, assumed to give the same contribution, in this homogenized continuum model.

Diffusion parameters in the model, which may depend on concentrations, temperature, pressure,  is hard to measure, and is often not accurately known \cite{rev2}. In the simplest case, it may be assumed to be constant, as a linearisation of the model about typical operating conditions and for small currents when potential drops in electrolyte and Ohmic losses may be ignored.

A linear model with constant diffusion coefficient is often still solved numerically, but admits an analytic solution, which is more accurate and fast, and can be used for testing numerical methods.

Despite simplicity, this still poses some challenges. For example, if battery if fully charged and concentrations in electrodes are uniform, and the current is instantaneously turned on, there is a clash between the initial and boundary conditions, however at any positive time, the solution is smooth, and in the limit as time goes to zero, solution approaches the initial condition everywhere, including the electrode boundary. The value of the solution at the boundary is most important, as it determines the voltage. This is still true of the initial condition is not uniform but is sufficiently smooth. The behavior of the solution in the short time limit is quite hard to accurately reproduce in the numerical approach.

One way to solve it is to perform the Laplace transform with respect to time, and then the inverse Laplace transform, see for example \cite{Khasin2019}. In this study, we report a different approach, based on Fokas universal transform method \cite{fz}, \cite{FokasB}. The method was initially developed for linear constant coefficient and integrable nonlinear PDEs in two dimensions, but may be modified to solve the spherically symmetric diffusion problem in 3 dimensions, with non-constant coefficients due to the use of spherical coordinates. The advantage of the method is that it represents a solution via Fourier-like contour integrals in a complex plane, with good control on analytic behavior of the integrand in the complex plane. This enables us to study the solution, for example its long and short time asymptotic. For example, deforming contour integrals appropriately, one may represent the solution as a series corresponding to the residues of the contour integration. Manipulating contour integrals in such a way, we can represent the solution as a series exponentially convergent for any positive time (with slower convergent terms explicitly computed using residues), and everywhere convergent at zero time. Apart from mathematical interest, study of asymptotic is important for the task of measuring the diffusion parameters, needed for the model to be predictive.

\section{Analytic Solution via Fokas method}
\bgea
\partial_{\tau} c = \frac{1}{r^2}\partial_r \lb D r^2 \partial_r c\rb ,  0 < r < R, \tau >0,
\label{eq:cR}
\enea
with initial condition ${n}_0 (r),$ and boundary conditions determined by prescribed inflow at $r=R$ (and regularity at $r=0$):
\bgea
c(r,0) ={c}_0 (r) , 0 \leq r \leq R;
\\
\ld \lb r^2 \partial_r c \rb \rv_{r\rightarrow 0^+} =0, \ld D \lb \partial_r c \rb \rv_{r=R} = \tilde{j} (\tau), \tau \geq 0.
\label{con:cR}
\enea
We consider here the case when diffusion coefficient in the electrode particle is a constant. (Such assumption is reasonable since there is a wide range of values for diffusion coefficients in experimental literature, and in practise diffusion coefficient is often not known apriori, rather it is deduced by fitting simulations to experimental data of interest).

To simplify subsequent considerations, we introduce dimensionless variables $x,t,n$
\bgea
r = R\  x , \tau = \frac{R^2 t}{D} ,    c = \frac{n}{R^3}.
\enea
In terms of those dimensionless variables, the equation and boundary conditions become:
\bgea
\partial_{t} n = \dst \frac{1}{x^2}\partial_x \lb x^2 \partial_x n\rb ,  0 < x < 1, t>0;
\label{eq:n}
\enea
\bgea
n(x,0) =\ni  (x) , 0 \leq x \leq 1;
\\
\ld \lb x^2 \partial_x n \rb \rv_{x\rightarrow 0^+} =0, \ld \lb \partial_x n \rb \rv_{r=R} = j (t), t\geq 0,
\label{con:n}
\enea
where $j(t) = \frac{R^4}{D} \ \tilde{j} (\frac{R^2 t}{D})$

\section{Solution via Fokas Method}

Most of the continuum modeling so far is numerical modeling. Analytical methods would complement those to (1) investigate solution (2) provide tests for numerical methods (3) serve as basis for {\it new} numerical methods. We illustrate this approach by a simple example of spherically-symmetric diffusion of Lithium ions into a spherical electrode particle.

Let $n= \frac{u}{x}$, then (\ref{eq:n}), (\ref{con:n}) become:
\bgea
\partial_t u =  \partial^2_x u, \quad 0 < x < 1, t >0,
\label{eq:u}
\enea
\bgea
u(x,0)= \ui(x)  , 0 \leq x \leq 1;
\\
\ld u \rv_{x \rightarrow 0^+} =0, \ld  \lb \partial_x u   - u \rb \rv_{x=1} = j(t), t > 0.
\label{con:u}
\enea
By Ehrenpreis's fundamental principle, made explicit by the unified transform method\cite{FokasB}, the solution of (\ref{eq:u})- (\ref{con:u}) can be given as a superposition of elementary solutions of  (\ref{eq:u})
\bgea
u(x, t) = \dst \int_{\la\in\Gamma} e^{  - \la^2 t + i \la x  }   d \mu(\la)
\enea
with an appropriate measure $\mu$ and integration contour $\Gamma \subset \Bbb{C}.$  This may be constructed as follows.

As observed in \cite{fz}, (\ref{eq:u}) is equivalent to the condition that the fundamental differential 1-form $\omega ,$
\bgea
\omega(x,t;k_0,k_1) = u e \ dx + \lb e \partial_x u - u \partial_x e \rb \ dt ,   e := \exp \lb -ik_0 t - i k_1 x \rb ,
\label{omega}
\enea
is closed whenever  the on-shell condition
\bgea
i k_0 +  k_1^2 =0
\label{eq:onshell}
\enea
is satisfied (such $k_0, k_1$ can be parameterized by  $k_1 = \la, i k_0 = -\la^2, \la\in \Bbb{C}$). Indeed,
\bgea
d \omega = \ls  e \lb u_{xx} - u_t\rb - u \lb e_{xx} + e_t\rb  \rs  dx\wedge dt,
\label{domega}
\enea
and $e_{xx} + e_t = -  \lb ik_0 + k_1^2 \rb e.$ As a consequence of (\ref{domega}), we can represent the solution by
\bgea
u(x,t) = \dst\int_{-\infty}^\infty  \int_{-\infty}^\infty \frac{dk_0 dk_1}{(2 \pi)^2} \frac{ e^{i k_0t + i k_1 x } }{ik_0+k_1^2} \dst\oint_{(x_1,t_1)\in\gamma}\omega(x_1,t_1;k_0,k_1) ,
\label{integral rep}
\enea
where $\gamma$ is a closed positively oriented contour in $D$  surrounding point $(x,t)$ once. Indeed, by Green's theorem (\ref{domega}) implies that
\bgea
\dst\oint_{(x_1,t_1)\in\gamma}\omega(x_1,t_1;k_0,k_1) = \int\int_{\Omega_{\ga}}dx_1 dt_1  u(x_1,t_1)(ik_0+k_1^2)e^{-i k_0 t_1 - i k_1 x_1 }
\enea
where $\Omega_{\ga}$ is a domain in $\Omega$ bounded by $\ga$, and the right-hand side in (\ref{integral rep}) is
\bgea
\dst\int_{-\infty}^\infty  \int_{-\infty}^\infty \frac{dk_0 dk_1}{(2 \pi)^2} \int\int_{\Omega_{\ga}}dx_1 dt_1 u(x_1,t_1) e^{ i k_0 (t-t_1) + i k_1 (x-x_1) } =
\\
 \int\int_{\Omega_{\ga}}dx_1 dt_1 u(x_1,t_1) \delta(x-x_1) \delta(t-t_1) = u(x,t).
\label{eq:delta}
\enea

As in \cite{fz}, we can perform one of the $k$ integrations by residues, using analytic  and decay at infinity properties of the integrand in (\ref{integral rep}), to get an integral representation of Ehrenpreis type, expressing the solution as a combination
of elementary solutions $e^{\imath \la x -\la^2 t}.$
\bgea
u(x,t) = \frac{1}{2 \pi} \lb \int_{-\infty}^\infty e^{\imath \la x -\la^2 t}\hui  (-\imath \la)d\la
- \int_{\partial \Dp} e^{\imath \la x -\la^2 t} \lb n_0 (\la^2,\tpg) + i \la d_0 (\la^2,\tpg) \rb d\la \rd
\\
\ld - \int_{\partial \Dm} e^{\imath \la (x-1) -\la^2 t}\lb   n_1 (\la^2,\tpg) + i \la d_1 (\la^2,\tpg)  \rb d\la \rb,
0<x<1, 0<t< \tp ,
\enea
where
\bgea
d_0 (\la,\tpg) = \dst\int_0^{\tpg} e^{\la t_1} u(0, t_1) dt_1, \ n_0 (\la,\tpg) = \dst\int_0^{\tpg} e^{\la t_1} u_x (0, t_1) dt_1,
\\
d_1 (\la,\tpg) = \dst\int_0^{\tpg} e^{\la t_1} u(1, t_1) dt_1, \  n_1 (\la,\tpg) = \dst\int_0^{\tpg} e^{\la t_1} u_x (1, t_1) dt_1,
\\
\hui (\la) = \dst\int_0^1  e^{\la x_1} u(x_1, 0) dx_1;
\enea
\bgea
\Dp := \lbr \ld \la\in \Bbb{C} \rv \frac{\pi}{4}\leq arg \la \leq \frac{3 \pi}{4} \rbr,
\\
\Dm := \lbr \ld \la\in \Bbb{C} \rv \frac{5 \pi}{4}\leq arg \la \leq \frac{7 \pi}{4} \rbr;
\enea

\bgea
\ga = \gai \cup \ga_0 \cup \ga_1 \cup \gatp,
\\
\gai =  \lbr (s,0)\lv 0\leq s \leq  1  \rd \rbr, \ga_1 =  \lbr (1,s)\lv 0\leq s \leq  \tp \rd \rbr,
\\
\gatp =  \lbr (1-s,t_+)\lv 0\leq s \leq  \tp \rd \rbr, \ga_0=  \lbr (0,\tp - s)\lv 0\leq s \leq  \tp \rd \rbr ,
\\
\tp > t.
\label{ga}
\enea
Our initial and boundary conditions are posed on $\gai , \ga_0, \ga_1 .$  Those conditions provide complete information about the 1-form $\omega$ on $\gai,$  partial information about $\omega$ on $\ga_0, \ga_1$ (with, in effect, half of data, given by Dirichlet or Robin boundary conditions, known, while the other half needs to be determined by solving the global relation, discussed below). We do not know the solution on the future time slice  -this should come about from solving the equation. Since this is an evolution equation, we expect that the future time slice data either is not needed (or can be found); moreover, it should not matter a great deal how exactly we close the integration contour $\ga$, provided that $(x,t)$  in (\ref{integral rep}) is inside $\ga$ (and thus (\ref{eq:delta}) holds).

Separating contributions of the 4 pieces of $\ga,$ (\ref{ga}) to the right-hand side in (\ref{integral rep}),
$$u = u_{\gai} + u_{\ga_0}+ u_{\ga_1}+ u_{\gatp}.$$
It is easy to see that $\gatp$ contribution to the right-hand side in (\ref{integral rep}) is indeed zero. Indeed, the $k_0$ dependence in $u_{\gatp}$ is of the form $\frac{e^{-i k_0 \epsilon } }{ik_0+k_1^2},$ and $k_0$ integration contour can be closed in the lower half $k_0$ plane; and since the expression is an analytic function of $k_0$ in the lower half plane, the $k_0$ integral is zero,
\bgea
u_{\gatp} (x,t) = 0.
\enea
A similar consideration shows that $k_0$ integrand in $u_{\gai}$ is of the form $\frac{e^{ i k_0 t}}{ik_0+k_1^2},$  $t>0;$ thus the integration contour can be closed in the upper half plane. Computing the closed contour integration over $k_0$ via the residue at $k_0= i k_1^2$ and renaming the integration variable $k_1$ by $\la$ we get:
\bgea
u_{\gai} (x,t) = \dst\int_{-\infty}^\infty  \frac{d\la}{ 2 \pi  } \int_0^1 dx_1  \ui (x_1) e^{- \la^2 t   + i \la (x-x_1) }
\label{intrep_0}
\enea
Computation of
$u_{\ga_1} (x,t)$ is largely similar but now since the integrand in (\ref{integral rep}) is of the form
$$
\frac{e^{ik_0 (t-t_1) +i k_1 (x-1) }\lb  u_x (1,t_1) + i k_1 u (1, t_1)\rb}{ik_0+k_1^2}
$$
with $x-1<0$ while $t-t_1$ can be both positive and negative, we perform the $k_1$ integration first, by completing the integral in the lower half $k_1$ plane and using residues. When $0<k_0<\infty,$ the residue is computed at  $k_1 = e^{-i \frac{\pi}{4}}\sqrt{k_0},$ while when $- \infty< k_0<0,$ the residue is computed at  $k_1 = e^{-i \frac{3 \pi}{4}}\sqrt{-k_0}.$ Those at-residue values of $k_1$ traverse a curve $\Gamma_-$ when $k_0$ change from $-\infty$ to $\infty$,  with $\Gamma_-$ a union of two rays,
$\Gamma_- = \lbr -s e^{-i \frac{3 \pi}{4}} , -\infty<s<0 \rbr \cup \lbr s e^{-i \frac{\pi}{4}} , 0<s<+\infty \rbr .$
We still need to integrate the result over $k_0$, and it is convenient to change integration variable to $\la \in \Gamma_-$, $k_0 = i \la^2,$ $\la \in \Gamma_-$ (that is, in effect, to go back to $k_1$ as the remaining $k$ space integration variable, using that at the residue  $k_0 = i k_1^2$ and $\frac{dk_0}{2k_1}= i dk_1.$) As a result we get
\bgea
u_{\ga_1} (x,t) = \dst\int_{\Gamma_-}  \frac{d\la}{ 2 \pi } \int_0^t dt_1  \lb u_x (1,t_1) + i \la u (1, t_1) \rb  e^{- \la^2 (t-t_1)   + i \la (x-1) }.
\label{intrep_iR}
\enea
A similar computation gives
\bgea
u_{\ga_0} (x,t) = \dst\int_{\Gamma_+}  \frac{d\la}{2 \pi} \int_0^t dt_1  \lb u_x (0,t_1) + i \la u (0, t_1) \rb  e^{- \la^2 (t-t_1)   + i \la x},
\\
\Gamma_+ =\lbr - s e^{i \frac{\pi}{4}} , -\infty<s<0 \rbr \cup \lbr s e^{ i \frac{3\pi}{4}}, 0<s<+\infty \rbr .
\label{intrep_i0}
\enea

\subsection{Global relation}
Integral representations (\ref{intrep_0})- (\ref{intrep_i0}) involve both known and unknown boundary data. This may be accomplished by supplementing those integral representations with boundary conditions and solving corresponding linear integral equations to find the missing data. However simpler way to do so may arise by solving the global relation of the unified transform method \cite{fg} -\cite{FokasB} . The latter arise by noting that when the on-shell condition (\ref{eq:onshell}) is satisfied, the 1-form $\omega$ is closed, and therefore, a line integral of $\omega$ over a closed contour is zero:
\bgea
(ik_0 +k_1^2 = 0) \  \Rightarrow \ \oint_\ga \omega = 0.
\label{eq:oint}
\enea
Using parametrization $k_1=\la, k_0 = i \la^2$,  (\ref{eq:oint}) implies the global relation between known and unknown initial and boundary data
\bgea
\hat{u}_0 (- i \la) + e^{- i \la} \lb n_1 (\la^2,\tpg) + i \la d_1 (\la^2,\tpg)  \rb - \lb n_0 (\la^2,\tpg) + i \la d_0 (\la^2,\tpg) \rb
\\
- \hat{u} (- i \la, \tpg)e^{\la^2 \tpg} = 0.
\label{eq:global}
\enea
Here $d_i,n_i, i =0,1$ are Fourier-like transforms of the Dirichlet, Neumann boundary data, respectively at $x=0, x=1$, and $\hat{u}_0, \hat{u}$ transforms of the initial and time $\tpg$ data:
\bgea
d_0 (\la,\tpg) = \dst\int_0^{\tpg} e^{\la t_1} u(0, t_1) dt_1, \ n_0 (\la,\tpg) = \dst\int_0^{\tpg} e^{\la t_1} u_x (0, t_1) dt_1,
\\
d_1 (\la,\tpg) = \dst\int_0^{\tpg} e^{\la t_1} u(1, t_1) dt_1, \  n_1 (\la,\tpg) = \dst\int_0^{\tpg} e^{\la t_1} u_x (1, t_1) dt_1,
\\
\hat{u} (\la, \tpg) = \dst\int_0^1  e^{\la x_1} u(x_1, \tpg) dx_1 , \ \hat{u}_0 (\la)=\hat{u} (\la, 0).
\enea
Our boundary conditions (\ref{con:u}) imply that
\bgea
d_0 (\la,\tpg) =0, d_1 (\la,\tpg) = n_1 (\la,\tpg) - \hat{j} (\la,\tpg),
\\
\hat{j} (\la,\tpg)= \dst\int_0^{\tpg} e^{\la t_1} j(t_1) dt_1.
\label{eq:bc_dn}
\enea
Using (\ref{eq:bc_dn}), the global relation becomes
\bgea
n_0 (\la^2,\tpg) - e^{-\imath \la} (1+\imath \la) n_1 (\la^2,\tpg) = G(-\la, \tpg),
\enea
where
\bgea
G(\la,\tpg):= g(\la,\tpg)   - \hat{u} ( i \la, \tpg)e^{\la^2 \tpg}  ,
\\
g(\la,\tpg): = \hat{u}_0 ( i \la) + i \la e^{i\la}\hat{j} (\la^2,\tpg),
\enea
Since $\hat{u}_0 (\la)$ is known from the initial conditions, there are two unknown functions, $n_0$ and $n_1$ in the global relation (as well as the future data $\hat{u} (\la, t),$ which we expect will eventually not contribute, as indeed will be  shown below). However, the $n_0 (\la^2,t)$ and $n_1 (\la^2,t)$ terms in the global relation are invariant under $\la\rightarrow -\la$ , while other terms in the global relation are not; this means that there is another independent global relation involving the same unknown functions, obtained by replacing $\la$ by $-\la.$   Introduce $h(\la)$  by
\bgea
h(\la) = e^{-\imath \la} (1+\imath \la)
\enea
(note that $h(\la)$  is related to spherical Hankel functions $h_n^{(2)} (z),$ \cite{AS}: $h(\la) = i \la^2 \ld \lb \frac{d}{dz} h_0^{(2)} (z)\rb\rd_{z=\la}= - i \la^2 h_1^{(2)}(\la)\ $ ). Supplementing (\ref{eq:global}) by an equation with $\la\rightarrow -\la$, we can solve those two relations for $n_0$ and $n_1,$ using just linear algebra:
\bgea
\lb \begin{array}{c} n_0 (\la^2,\tpg) \\n_1 (\la^2,\tpg)  \ena\rb = \dst \frac{1}{\Delta(\la) }
\lb \begin{array}{cc} - h(-\la)  & h(\la)  \\
-1 & 1
 \ena \rb
\lb \begin{array}{c} G(-\la,\tpg) \\ G(\la ,\tpg)  \ena\rb,
\\[4mm]
\Delta(\la) =  h(\la)- h(-\la).
\label{eq:n0n1}
\enea
Let us separate contributions of the initial and  boundary conditions $g (\pm \la,\tpg)$ and the contribution of future-time
slice, $\hat{u}   ( \pm \imath \la,\tpg)e^{\la^2 \tpg}$:
\bgea
n_0 (\la^2,\tpg) = \dst\frac{\nu_0 (\la) + \nu_0^{+} (\la)}{\Delta(\la) };
\\
\nu_0 (\la)  = h(\la) g (\la,\tpg)  - h(-\la) g (-\la,\tpg)
\\
\nu_0^{+} (\la)  = - \lb h(\la) \hat{u}   (\imath \la, \tpg) - h(-\la) \hat{u}  (-\imath \la,\tpg)  \rb e^{\la^2 \tpg}
\\[2mm]
n_1 (\la^2,\tpg) = \dst\frac{\nu_1 (\la) + \nu_1^{+} (\la)}{\Delta(\la) };
\\
\nu_1(\la)=  g (\la,\tpg)- g (-\la,\tpg)
\\
\nu_1^{+}(\la) = - \lb   \hat{u}   (\imath \la, \tpg) -  \hat{u}  (-\imath \la,\tpg)  \rb e^{\la^2 \tpg}
\enea

The future time data do not contribute to the solution, as the integrand is analytic and decays at infinity, respectively in $\Dp$ and $\Dm .$ We then have representation of the solution in the Ehrenpreis form,
\bgea
u(x,t) = \frac{1}{2 \pi} \lbr  \int_{-\infty}^\infty e^{\imath \la x -\la^2 t}\hat{u}_0  (-\imath \la)d\la

- \int_{\partial \Dp} \frac{e^{\imath \la x -\la^2 t}}{\Delta(\la)}  \mu_0 (\la)  d\la
- \int_{\partial \Dm} \frac{e^{\imath \la x -\la^2 t}}{\Delta(\la)} \mu_1 (\la) d\la
\rbr
\\
\mu_0 (\la) =
\lb \imath \la +1 \rb e^{-\imath \la} \hat{u}_0 ( i \la)  +
 \lb \imath \la -1 \rb e^{\imath \la} \hat{u}_0 ( -i \la) + 2 \imath \la \hat{j} (\la^2,t),
 \\
\mu_1 (\la) =
\mu_0 (\la) -  \Delta(\la) \hat{u}_0  (-\imath \la) ,
\\
\Delta(\la) =  e^{-\imath \la} (1+\imath \la)- e^{\imath \la} (1-\imath \la),
\\
\Dp := \lbr \ld \la+ \imath \veps \in \Bbb{C} \rv \frac{\pi}{4}\leq arg \la \leq \frac{3 \pi}{4} \rbr,
\\
\Dm := \lbr \ld \la - \imath \veps\in \Bbb{C} \rv \frac{5 \pi}{4}\leq arg \la \leq \frac{7 \pi}{4} \rbr,
\\
\hat{j}(\la, \tpg) = \int_0^{\tpg} e^{\la \tau} j(\tau) d\tau ,
\\[1mm]
\hat{u}_0  (\la) = \int_0^1 e^{\la x} {u}_0  (x) dx.
\label{sol:integral}
\enea
We note that integration contours in the three integrals above are different; would they be the same, such an expression would be zero, due to the global relation. This implies that for sufficiently regular initial and boundary conditions the solution may be computed by using residues, for example by deforming all integration contours towards the real axes (provided there is sufficient decay at infinity, which would be the case for positive times)

\begin{lem}
(1) All zeroes of $\Delta( \la)$ are on the real line  $Im \ \la =0$;(2)  $\Delta( \la)$ has a zero of order 3 at $\la= 0,$ and $\Delta( \la)= \frac{-2 i \la^3}{3} + O(\la^5), \la \rightarrow 0,$  moreover  if $\la_k\neq  0$ is a zero of $\Delta( \la),$ then $-\la_k$ is also a zero of $\Delta( \la).$
\end{lem}
{\it Proof}.  (1) follows from the fact that zeroes of $\Delta( \la)$ are eigenvalues of a self-adjoint operator, the Laplacian, with the appropriate boundary conditions, and therefore all the eigenvalues must be on the real line; (2) is explicit.

Noted: one may also approach this by using the maximum principle, however it is more complicated that way to arrive at the conclusion.

\section{Study of the solution}
\subsection{Examples}
\subsubsection{Charging or discharging by a constant current from uniform concentration}
We will illustrate here behavior of the solution by an example when initial concentration is uniform (this corresponds to the initial condition $u_0(x)$ being linear in $x$, $u_0(x) = c_0 x $ see (\ref{eq:u}) and above),   and a constant flux $j(t) =j_0$ is turned on at time equals zero.

We have in that case that $\hat{j}(\la^2, t ) = \frac{e^{\la^2 t}-1}{\la^2} j_0;$ however the term with $e^{\la^2 t}$ does not contribute to the solution, as it gives an expression which decays at the infinity in $\Dp$ and $\Dm$; it is therefore sufficient to replace $\hat{j}(\la^2, t )$ by $\frac{-1}{\la^2} j_0.$ We also have that $\hat{u}_0 ( \imath \la)= \frac{e^{\imath \la} (\imath \la -1) +1}{\la^2}.$  As noted above, the integration can be performed by deforming onto the real axes and using residues. The contribution of the uniform initial condition is only at the pole  $\la=0;$ it therefore remains unchanged and equals $c_0 x.$ Contribution of the current term from the residue at $\la =0$ is $x \lb 3 t + \frac{5x^2 -3 }{10} \rb j_0$. Contribution of nonzero poles on the real axes, combining $\la$ and $-\la$ poles, is
$$
S = -2 j_0 \dst\sum_{\la  \in \mbox{\tiny BesselJZero $(\frac{3}{2})$} } \frac{\sin (\la x)}{\sin \la} \frac{e^{-\la^2 t}}{\la^2},
$$
where BesselJZero $(\frac{3}{2})$ are (positive) zeroes of the BesselJ $(\frac{3}{2})$ function.

Combining,
\bgea
u(x,t) = c_0 x + x \lb 3 t + \frac{5x^2 -3 }{10} \rb j_0 -2 j_0 \dst\sum_{\la  \in \mbox{\tiny BesselJZero}(\frac{3}{2})} \frac{\sin (\la x)}{\sin \la} \frac{e^{-\la^2 t}}{\la^2},
\label{sol}
\enea

Large positive zeroes of the BesselJ $(\frac{3}{2})$ are approximate zeroes of $\cos\la$, and are approximately $\frac{\pi}{2} n, n\in \Bbb{N}.$  Therefore the series can be majorated by $\dst\sum_{n\in \Bbb{N} }  \frac{1} {n^2},$ is absolutely and uniformly convergent, and define a continuous function for all $t\geq 0, 0\leq x\leq 1,$ and a smooth function for $t>0.$ Manipulating with contour integration similar to the above, that function equals the initial condition $c x$ when $t=0.$ Morevover it is explicit that the boundary condition is satisfied for all positive $t$. Thus (\ref{sol}) is the classical solution of (\ref{eq:u}, \ref{con:u}). This corresponds to concentrations $n(x,t),$ (\ref{eq:n} - \ref{con:u}),
\bgea
n(x,t) =\frac{u(x,t)}{x},
 \enea
and radial linear density $4\pi x u(x,t).$

We note that computation of zeroes of the Bessel functions (investigated since Dr. G.N. Watson in 1918) are coded numerically, for example in Mathematica, which makes it easy to implement (given a Mathematica). An alternative to the series may be to compute the integrals numerically. As the integrand is analytic, Gaussian quadratures may be used. To achieve better convergence at positive times, the integration contours may be deformed for better decay as $\la \rightarrow \infty.$
\subsubsection{Polynomial in time current, smooth initial condition}
If the current at the boundary $j(t)$ is a polynomial in time  $t$ (in applications such polynomial may be an approximation of the actual current), then since the problem is linear it is sufficient to consider monomials, $j_n(t) = t^n, n= 1,2,\ldots$ We compute the transform $\hat{j} (\la^2, t)$; and similarly to the constant case, the terms with $e^{\la^2 t}$ in $\hat{j} (\la^2, t)$ do not contribute to the solution as the integrand decays as $\la\rightarrow \infty$ in  $\Dp, \Dm .$ Thus it is sufficient to take the term not containing $e^{\la^2 t},$ that is
\bgea
\dst \hat{j}_n = \frac{n!}{(-\la^2)^{1+n}}
\enea
Similarly to the above, contribution of such current to the solution, assuming zero initial condition,  can be expressed as exponentially convergent series
\bgea
\dst - 2 res_{\mbox{\tiny $\la =0$}} \lb \frac{(-1)^{1+n} n! e^{-\la^2 t + \imath \la x}}{\la^{2n+1} \Delta (\la) }\rb +
\dst n! \sum_{\la  \in \mbox{\tiny BesselJZero}(\frac{3}{2})} \frac{\sin (\la x) }{\sin \la}  \frac{e^{-\la^2 t}}{(-\la^2)^{1+n}}.
\label{cont_curr}
\enea
It is easy to see that (\ref{cont_curr}) satisfy for $t>0$  the boundary conditions. Indeed,  the terms in the series satisfy the boundary conditions with zero current, as it is enough to verify term by term, using the fact that $\la$ are zeroes of the BesselJ$(\frac{3}{2})$ function. For the residue term, one can start with verifying boundary conditions for the constant current, corresponding to $n=0$  (and this is already done in section 4.1.1), and integrating those conditions with respect to time $n$ times. The terms in the residue which do not contain the $e^{-\la^2 t}$ factor will not contribute, as in terms of contour integration, the integrand is analytic and decays fast enough (as $\frac{1}{\la^2}$ or better) at infinity, and integration contour can be deformed to infinity, providing zero contribution.  

Assuming that initial condition is smooth, contribution of the initial condition to the solution, at zero current at the boundary, is
\bgea
3 x \int_0^1 x_1 \ui (x_1) dx_1  - 
\\
-\sum_{\la \in \mbox{\tiny BesselJZero}(\frac{3}{2})} \frac{\imath e^{-\la^2 t} \sin (\la x) }{\la \sin \la}  
\lb  e^{-\imath \la} (1+\imath \la)  \hui(\imath \la ) - e^{\imath \la} (1-\imath \la)   \hui(-\imath \la ) \rb 
\enea 
(the term $3 x \int_0^1 x_1 \ui (x_1) dx_1 = 3 x \frac{d}{d\la} \ld \hui \rv_{\la=0} $ appears from computing the residue at $\la=0$).
\subsection{Convergence to the initial condition}
For simplicity, assume that the current $j(t)$ has a piecewise continuous derivative, and the initial condition $\ui(x)$ as well as its first and second derivatives are continuous for all $x,$ $0\leq x\leq 1.$ Then as $t\downarrow 0,$ $u(x,t) \rightarrow \ui(x).$ 

To show this,  we describe below contributions of various terms by listing their contributions to the integrand in the integration over  $\partial D_+$ in the integral representation (\ref{sol:integral}), taking into account that this integrand equals to the sum of integrands over $\partial \Dm$ and the real axes, with integration contour for the former oriented oppositive to the latter two integration contours. As a result, computation of integrals  can be reduced to an appropriate computation of residues, as specified below.
 
Integrating by parts we may assume that WLOG
\bgea
\hat{j} (\la^2,t) = \frac{1}{\la^2} \eta(\la^2,t), \quad \eta(\la^2,t) : = j(t) e^{\la^2 t} - j(0) - \hat{j^\prime} (\la^2,t)
\enea
Thus contribution of $\hat{j} (\la^2,t)$  tends to zero uniformly in $x$  as $t\downarrow 0,$ as the integrand is bounded, and is in fact majorated by $\frac{C(t)}{\la^2}$ with $C(t) \rightarrow 0 $ (in particular since the $t$ integral defining the transform $\hat{j^\prime} (\la^2,t)$ is over the t-interval of length $\lv t \rv,$ while the integrand is bounded as a function of $t, \la$).

As for the contribution of the initial condition, integrating by parts and using $\ui(0) =0$, we have that 
\bgea
\dst \hui (\la) = \frac{ \ui (1) e^\la}{\la}  -   \frac{\ui^\prime (1) e^{\la}-\ui^\prime (0)}{\la^2} + \frac{\hui^{\prime\prime}(\la)}{\la^2} 
\label{cont_init}
\enea
Contribution of the first term is
\bgea
\dst \frac{2 \ui (1) e^{-\la^2 t + \imath \la x}}{\imath \la \Delta (\la)}
\label{cont_init_c}
\enea
Since $\frac{\la  e^{ \imath \la x} } {\Delta (\la)} $ is bounded in both upper and lower half planes, the integrand is majorated by $\frac{1} {\lv \la\rv^2}$ on $\partial \Dp, \partial \Dm$ uniformly in $x,t$  and at $t=0$ is analytic in $  \Dp,   \Dm .$ Thus at $t=0$ contribution of this term is zero, as we can deform the integration contour to infinity, and as $t\downarrow 0$ it converges to 0 uniformly.

Contribution of the second term in (\ref{cont_init}) is 
\bgea
\dst - e^{-\la^2 t + \imath \la x} \lb \frac{\ui^\prime (0)}{\la^2} + \frac{2  \ui^\prime (1)} {\imath \la \Delta (\la)} \rb
\label{pr}
\enea
Contribution of the first term in (\ref{pr}) is the residue at $\la=0,$ that  is 
\bgea
x \ui^\prime (0),
\label{contr1r}
\enea
while contribution of the second term vanishes as $t\rightarrow 0,$ for the same reason as contribution of (\ref{cont_init_c}), as the residue at infinity vanishes. 

Contribution of the last term when $t>0$ is 
\bgea
\dst e^{-\la^2 t+ \imath x} \lb \frac{e^{\imath \la} (1-\imath \la) }{\la^2 \Delta(\la)} \hui^{\prime\prime}(-\imath \la) - \frac{e^{-\imath \la} (1+\imath \la) }{\la^2 \Delta(\la)}\hui^{\prime\prime}(\imath \la) \rb, 
\enea
and is majorated by $\frac{1}{\lv \la\rv^2}$ on the integration contour which may be taken as running above and below the real axes, by deforming integration contours in (\ref{sol:integral}) accordingly. It is therefore clear that this contribution is continuous in $t$, and to study the $t \downarrow 0$ limit we may take $t=0$. When $t=0$, since $\hui^{\prime\prime}(\imath \la) = \int_0^1 e^{\imath \la x_1} \ui^{\prime\prime}  (x_1) dx_1, $ we see using the explicit form of $\Delta(\la)$ in  (\ref{sol:integral}) that contribution of  $\hui^{\prime\prime}(\imath \la)$ decays as $\frac{1}{\la^2}$ or faster in both $\Dm$ and $\Dp,$ and therefore the contribution of this term is zero, as we can deform integration contours to infinity. Now writing 
\bgea
\hui^{\prime\prime}(-\imath \la)  = \huim^{\prime\prime}  (-\imath \la) + \huip^{\prime\prime} (-\imath \la), 
\\
\huim^{\prime\prime}  (-\imath \la) = \int_0^x e^{-\imath \la x_1} \ui^{\prime\prime}  (x_1) dx_1, \huip^{\prime\prime}  (-\imath \la) = \int_x^1 e^{-\imath \la x_1} \ui^{\prime\prime}  (x_1) dx_1,
\enea
we see that  the $\huip^{\prime\prime} (-\imath \la)$ term decays at infinity as well, and the contribution of that term is zero when $t=0.$ We are thus left with computing the contribution of
\bgea
e^{\imath \la x}  \frac{e^{\imath \la } (1-\imath \la) }{\la^2 \Delta(\la)} \huim^{\prime\prime}(-\imath \la). 
\enea
Since $\Delta(\la) = e^{-\imath \la} (1+\imath \la)- e^{\imath \la} (1-\imath \la),$ adding and subtracting $e^{-\imath \la} (1+\imath \la)$ in the numerator this equals
\bgea
  \frac{-e^{\imath \la x} }{\la^2 } \huim^{\prime\prime}(-\imath \la) + e^{\imath \la x}  \frac{e^{-\imath \la } (1+\imath \la) }{\la^2 \Delta(\la)} \huim^{\prime\prime}(-\imath \la)
  \label{surge}
\enea
Similarly to the above, the second term in (\ref{surge}) is decaying at infinity and does not contribute, while contribution of the first term is the residue at $\la =0$. Computing the residue, we get
\bgea
x \huim^{\prime\prime}(0)  - \frac{d}{d\la} \ld \huim^{\prime\prime}\rv_{\la=0}.
\label{contr2mr}
\enea
But it follows from definition that 
\bgea
\huim^{\prime\prime}(0) = \int_0^x \ui^{\prime\prime} (x_1) dx_1= \ui^\prime (x) - \ui^\prime (0),
\\
\frac{d}{d\la} \ld \huim^{\prime\prime}\rv_{\la=0} = \int_0^x x_1 \ui^{\prime\prime} (x_1) dx_1  = x \ui^\prime (x) - \ui(x) + \ui (0).
\label{contr2m}
\enea
Combining contributions (\ref{contr1r}), (\ref{contr2m}), (\ref{contr2mr}), and using $\ui(0)=0, $  we get the desired result that $u(x,t) \rightarrow \ui(x)$ as $t\downarrow 0$.
\section{Acknowledgment}This research was supported by the Faraday Challenge Grant - Multiscale Modeling project.

Discussions with Michael Khasin, Karen Martirosyan, Charles Monroe, Thanasis Fokas, Irina Starikova are much  appreciated.

Noted: results reported here were obtained and internally reported in February 2020.
Further study of the solution and the final write-up is ongoing and will be updated.

\end{document}